\documentclass[aps,prl,superscriptaddress,reprint,twocolumn,longbibliography,amsmath,amssymb]{revtex4-1}
\usepackage{graphicx}  
\usepackage{epstopdf}
\usepackage{dcolumn}   
\usepackage{bm}        
\usepackage{amsmath}
\usepackage{verbatim}   
\usepackage{setspace}

\begin{document}

\title{Chiral magnetoresistance in Pt/Co/Pt zigzag wires}
\author{Yuxiang Yin}
\email[E-mail: ]{y.yin@tue.nl}
\author{Dong-Soo Han}
\author{June-Seo Kim}
\author{Reinoud Lavrijsen}
\affiliation{
  Department of Applied Physics, Center for NanoMaterials, Eindhoven University of Technology, PO Box 513, 5600 MB Eindhoven, The Netherlands
   }
\author{Kyung-Jin Lee}
\affiliation{
Department of Materials Science and Engineering, Korea University, Seoul 02841, Korea
   }
   \affiliation{
KU-KIST Graduate School of Converging Science and Technology, Korea University, Seoul 02841, Korea
   }
\author{Seo-Won Lee}
\affiliation{
Department of Materials Science and Engineering, Korea University, Seoul 02841, Korea
   }
\author{Kyoung-Whan Kim}
\affiliation{
  Center for Nanoscale Science and Technology, National Institute of Standards and Technology, Gaithersburg, Maryland 20899, USA
   }
\author{Hyun-Woo Lee}
\affiliation{
PCTP and Department of Physics, Pohang University of Science and Technology, Pohang 37673, Korea
}
\author{Henk J. M. Swagten}
\author{Bert Koopmans}
\affiliation{
  Department of Applied Physics, Center for NanoMaterials, Eindhoven University of Technology, PO Box 513, 5600 MB Eindhoven, The Netherlands
   }

\begin{abstract}
The Rashba effect leads to a chiral precession of the spins of moving electrons while the Dzyaloshinskii-Moriya interaction (DMI) generates preference towards a chiral
profile of local spins. We predict that the exchange interaction between these two spin systems results in a `chiral'
magnetoresistance depending on the chirality of the local spin texture. We observe this magnetoresistance by measuring the domain wall (DW)
resistance in a uniquely designed Pt/Co/Pt zigzag wire, and by changing the chirality of the DW with applying an in-plane magnetic field. A
chirality-dependent DW resistance is found, and a quantitative analysis shows a good agreement with a theory based on the Rashba model.
Moreover, the DW resistance measurement allows us to independently determine the strength of the Rashba effect and the DMI simultaneously, and the result implies a possible correlation between the Rashba effect, the DMI, and the symmetric Heisenberg exchange.
 \end{abstract}

\maketitle
\begin{figure}
\includegraphics[scale=0.135]{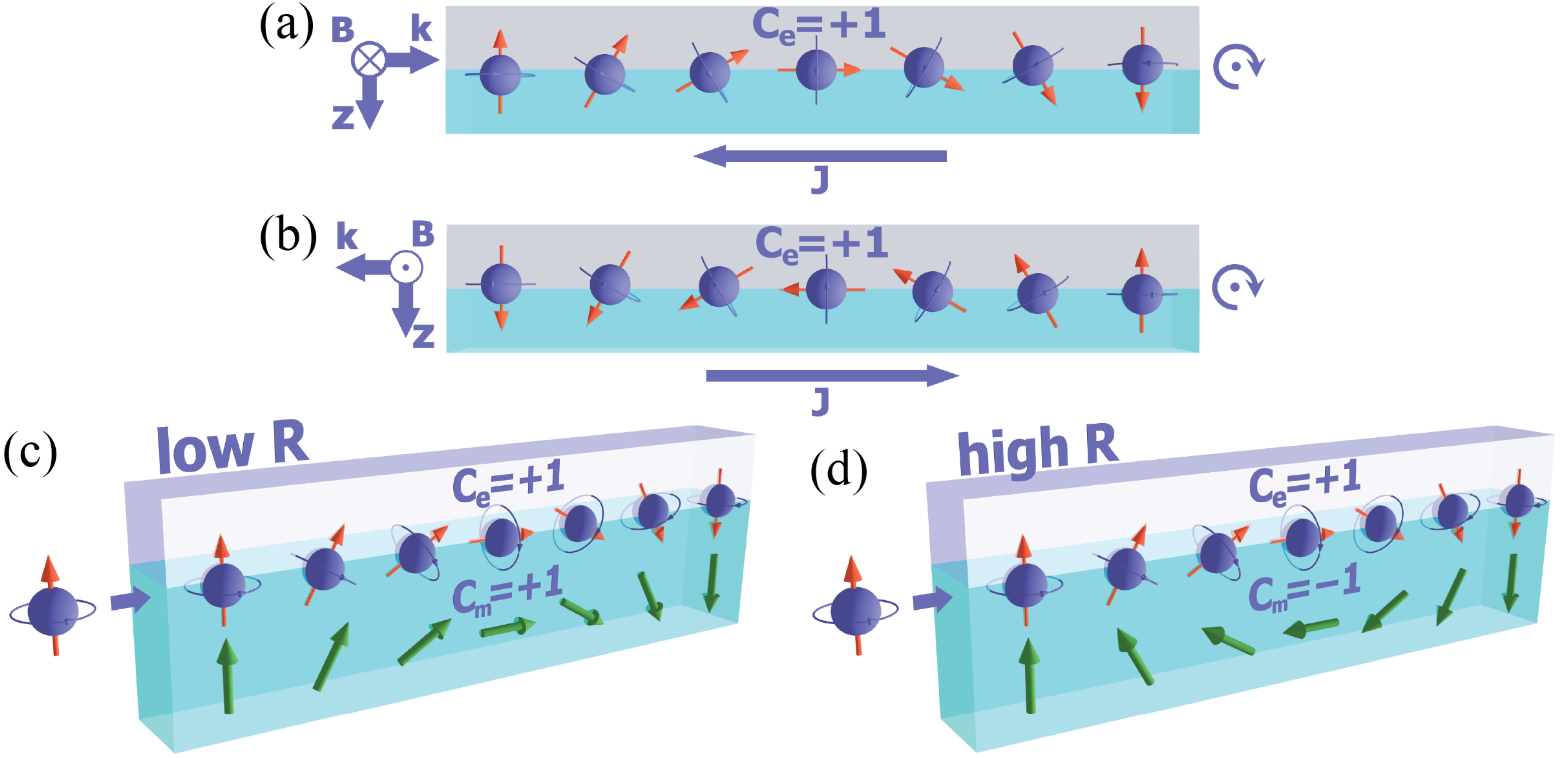}
\caption{\label{Figure1}(a)(b) Spin configurations of electrons traveling with opposite directions ($k>0$ or $k<0$) in the presence of a Rashba field $B$ at an interface ($z$ is interface normal). The chirality of the  profile of the spin precession is identical ($C_\textrm{e}=+1$) for both signs of $k$. (c)(d) Inside a chiral magnetic texture with chirality $C_\textrm{m}$ (green arrows), the rotation of electron spins generally follow the local spins due to exchange interaction while finite degree of misalignment is inevitable, which gives rise to the DW resistance. The resistance is lower (higher) when the misalignment is suppressed (enhanced) by the Rashba effect. Please note that the misalignment is very much exaggerated in (d) for better legibility.}
\vspace{-1.5em}
\end{figure}

In a magnetic system with inversion symmetry breaking combined with
spin-orbit coupling (SOC), a variety of chirality-related phenomena occur. For instance, due to the
Rashba effect \cite{bychkov1984properties}, the spins of the conduction electrons flowing at an interface are subject to an effective in-plane
magnetic field due to the relativistic SOC, resulting in spin precession around the field during its transport. In this case, the direction of the magnetic field depends on the electron flow direction. In contrast, the precession of the conduction spins, as can be seen from comparing
Figs.~\ref{Figure1}(a) and \ref{Figure1}(b), is independent of the flow direction and shares the same rotational sense (denoted by a chirality of electrons, $C_\textrm{e}$) \cite{Lounis2012}. Apart from
the chiral behavior of the conduction spins, the chiral nature of localized spins has recently been discovered in ferromagnetic materials with
inversion asymmetry and SOC. This gives rise to the Dzyaloshinskii-Moriya interaction (DMI) \cite{dzyaloshinsky1958thermodynamic,Moriya1960} leading to a chiral spin texture of the localized spins ($C_\textrm{m}$), manifested as N\'eel type magnetic domain walls (DWs) \cite{Chen2013} and magnetic skyrmions \cite{Woo2016}, which are crucial to the future design of spintronic devices.

Although these two effects originate in different spin systems, one can speculate about their interplay through exchange interaction between the
conduction and localized spins \cite{Jue2015,Kim2013,Pyatakov2014}. This results in a magnetoresistance (MR) which arises when the
conduction electron spins propagate with a fixed chirality due to Rashba-type SOC and interact with the chiral DMI-induced magnetic
texture. As shown in Figs.~\ref{Figure1}(c) and \ref{Figure1}(d), this should lead to lower (higher) resistance when $C_\textrm{e}$ is
identical (opposite) to $C_\textrm{m}$. This MR can be termed as `chiral MR', since the magnitude of the MR varies for
spin textures with different chiralities. However, a direct experimental evidence of this chiral MR is not found so far. Therefore, the
measurement of chiral MR could not only enrich the spectrum of chirality-related physical phenomena but also shed light on the
debated roles of the DMI and Rashba effect for fast current-driven DW motion \cite{Miron2011b} and other SOC-related
phenomena \cite{Kobs2011,Nakayama2013}.  Besides, this type of magnetoresistance provides further insights in engineering magnetic thin film systems with specifically desired chiral entities for future applications in nanoelectronics.

In this work, a chiral MR is observed by measuring the magnetic DW resistance, which probes the interaction between an electric current and a
magnetic DW. The magnetic DW is a
twisted spin structure whose chirality can be altered by an external field \cite{Ryu2013}, making it a perfect playground for measuring the
chiral MR. We report results on a controllable, specifically designed DW system where chiral DWs are created by modifying the perpendicular magnetic anisotropy of a zigzag wire using a focused ion beam, and applying an
in-plane field for switching between
the DW chiralities. It is found that the DW resistance varies with the chirality of DWs, and the observed behavior can be described by a theoretical model based on the Rashba Hamiltonian, lending strong support to our interpretation of a chirality-dependent interaction between the conduction and localized spins.
\begin{figure}
\includegraphics[scale=0.55]{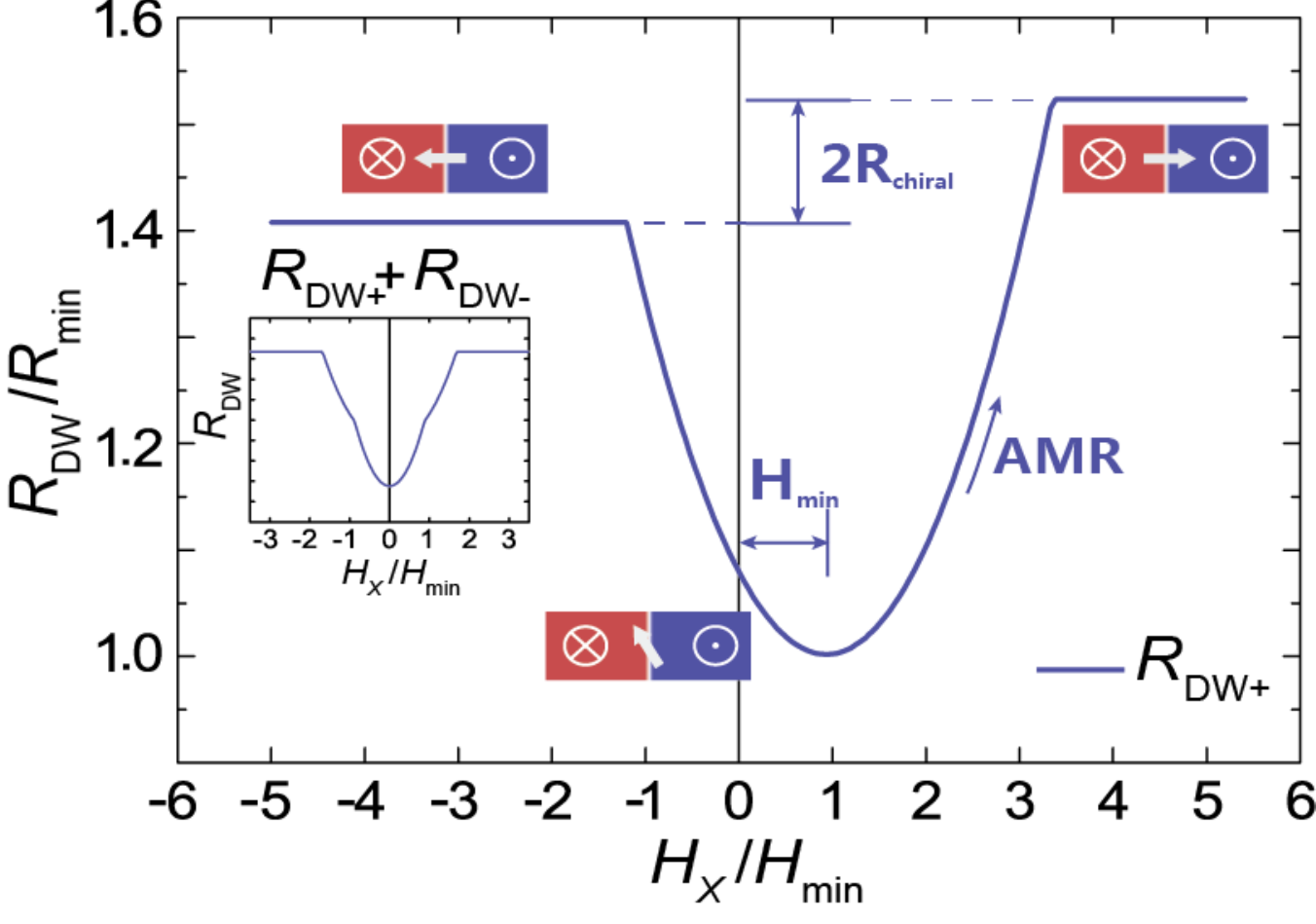}
\caption{Theoretical prediction of $R_{\textrm{DW+}}/R_{\textrm{min}}$, where $R_{\textrm{min}}$ is the resistance at $H_\textrm{min}$, as a function of in-plane field $H_x$. The white arrows indicate the DW angle. Inset: The
total DW resistance for two DWs with opposite chiralities.
}
\label{Figure2}
\vspace{-1.5em}
\end{figure}

We first analytically derive the chiral MR explained in Fig.~\ref{Figure1}. For better legibility, we exaggerate the effect of Rashba SOC but mostly ignore the effect of exchange interaction between the conduction electrons and local magnetization. However, in real systems, the exchange interaction is much higher than the Rashba interaction. Thus we adopt this regime and start from the Rashba Hamiltonian $(\hbar^2k_\textrm{R}/2m_\textrm{e}){\boldsymbol{\sigma}}\cdot ({\bf{k}} \times {\bf{\hat z}})$ added to a strong exchange interaction $J{\boldsymbol{\sigma}}\cdot \bf{m}$. Here $k_\textrm{R}$ characterizes the preferred spin precession rate, $\hbar$ the planck constant, $m_\textrm{e}$ the electron mass, $\bf{k}$ the electron momentum, $\bf{\hat z}$ the interface normal, and $\boldsymbol{\sigma}$ the spin Pauli matrix, $J$ the exchange energy, and $\bf{m}$ the direction of magnetization. The Rashba model captures core effects of the strong SOC combined with inversion symmetry breaking at the interface \cite{park2013orbital}. Recently, it was predicted \cite{Kim2013} that linear effects of the chirality can be obtained from non-chiral theories simply by replacing $\partial_x \bf{m}$, where $x$ is the current flow direction, with the so-called chiral derivative ${\tilde \partial _x}{\bf{m}} = {\partial _x}{\bf{m}} + {k_{\rm{R}}}\left( {{\bf{\hat z}} \times {\bf{\hat x}}} \right) \times {\bf{m}}$. Here, we apply the chiral derivative to the DW resistance derived by Levy and Zhang \cite{Levy1997}, ${R_{{\rm{LZ}}}} = \rho_\textrm{DW} \mathop \smallint \nolimits {\left( {{\partial _x}{\bf{m}}} \right)^2}dx$, where $\rho_\textrm{DW}$ is the DW resistivity. By assuming a DW with Walker profile \cite{schryer1974motion} and after some algebra (see Supplementary Material), we obtain ${\tilde R_{{\rm{LZ}}}} = 2\rho_\textrm{DW} \left( {{\lambda ^{ - 1}} \pm {k_{\rm{R}}}\pi \sin \phi } \right)$, where $\lambda$ is the DW width, $\phi$ the DW in-plane angle, and $\mp$ refers to up-down and down-up DW's, respectively. The DW is a Bloch wall when $\phi=0$ and a N\'eel wall when $\phi=\pm\pi/2$. The second term in $\tilde R_{\textrm{LZ}}$ is the chiral DW resistance, which changes its sign when the DW chirality changes.

Apart from the contribution from the inhomogeneity of the magnetic texture, the resistance also depends on the local magnetization direction itself due to anisotropic magnetoresistance (AMR) \cite{Kobs2011}. Based on AMR, the resistance is lower (higher) where the magnetization direction is perpendicular (parallel) to the current flowing direction. The theory of AMR implies that the MR is proportional to ${\sin ^2}\phi $. After adding the AMR contribution to $\tilde{R}_{\rm{LZ}}$, the total DW resistance can be written as

\vspace{-1em}
\begin{equation}
R_{\textrm{DW}\pm} = \Delta {R_{{\rm{AMR}}}}{\sin ^2}\phi  + 2\rho_\textrm{DW} {\lambda ^{ - 1}} \pm {R_{{\rm{chiral}}}}\sin \phi.
\label{Eq2}
\end{equation}
where $\Delta {R_{{\rm{AMR}}}}$ is a proportionality constant of AMR and ${R_{{\rm{chiral}}}} = 2\rho_\textrm{DW} {k_{\rm{R}}}\pi$ is the chiral DW resistance.

Experimentally, $\phi$ can be tuned by applying an in-plane field $H_x$ which is perpendicular to the DW. To determine the relation between $\phi$ and $H_x$, we start from the energy functional $E = \mathop \smallint \nolimits[A{{\left( {{\partial _x}{\bf{m}}} \right)}^2} + {\mu _0}{H_\textrm{k}}{M_{\rm{S}}}\left( {1 - m_z^2} \right) +\allowbreak {\mu _0}{H_{{\textrm{d}}}}{M_{\rm{S}}}m_x^2 + D{\bf{\hat y}} \cdot \left( {{\bf{m}} \times {\partial _x}{\bf{m}}} \right) - {\mu _0}{H_x}{M_{\rm{S}}}{m_x}] dx$ \cite{thiaville2012dynamics}, where $A$ is the exchange stiffness, $H_\textrm{k}$ the perpendicular anisotropy, $H_\textrm{d}$ the demagnetizing field, $D$ the DMI parameter, and  $M_\textrm{S}$ the saturation magnetization. Applying the Walker ansatz reduces the energy functional to a function of $\phi$ only. The energy minimizing condition $\partial E/\partial \phi  = 0$ gives
\begin{equation}
 \sin \phi  = \frac{\pi }{2}\frac{{ \pm D + \mu_0{H_x}{M_\textrm{S}}\lambda }}{{{\mu_0H_{\textrm{d}}}{M_\textrm{S}}\lambda }},
\label{Eq3}
\end{equation}
Here, we ignore the dependence of $\lambda$ on $H_x$, which is too weak to affect the analysis (see Supplementary Material). By substituting Eq.~(\ref{Eq3}) into
Eq.~(\ref{Eq2}), we plot in Fig.~\ref{Figure2} the resulting $R_{\textrm{DW}}/R_{\textrm{min}}$ as a function of $H_x/H_{\textrm{min}}$, where $R_{\textrm{min}}$ and $H_{\textrm{min}}$ are the values where the DW resistance is minimal. Three observations can be made: (i) There is an increase of the resistance when $H_x$ increases. This is expected since the field
pulls the DW into a N\'eel state, thereby increasing the AMR. (ii) For a finite DMI, the DW is in
between Bloch and N\'eel type for $H_x = 0$, i.e., $0<\phi<\pi/2$ \cite{Franken2014a,Emori2014}. This tilt angle depends
on the magnitude of the DMI, allowing us to determine the DMI energy density $D = \mu_0 H_{\textrm{DMI}} M_\textrm{S}\lambda$, where $H_{\textrm{DMI}}$ is the effective DMI field, by measuring the field shift
$H_{\textrm{min}}$ (see Supplementary Material). (iii) The most striking feature from this theory is a difference in $R_{\textrm{DW}}$ (indicated by $2R_\textrm{chiral}$ in Fig.~\ref{Figure2}) at high applied in-plain fields showing the unique dependence of the DW resistance on chirality. It is therefore very useful to measure the in-plane field dependence of DW
resistance, since it allows us to independently determine the strength of DMI and the Rashba effect simultaneously (see Supplementary Material).

\begin{figure}
\includegraphics[scale=0.25]{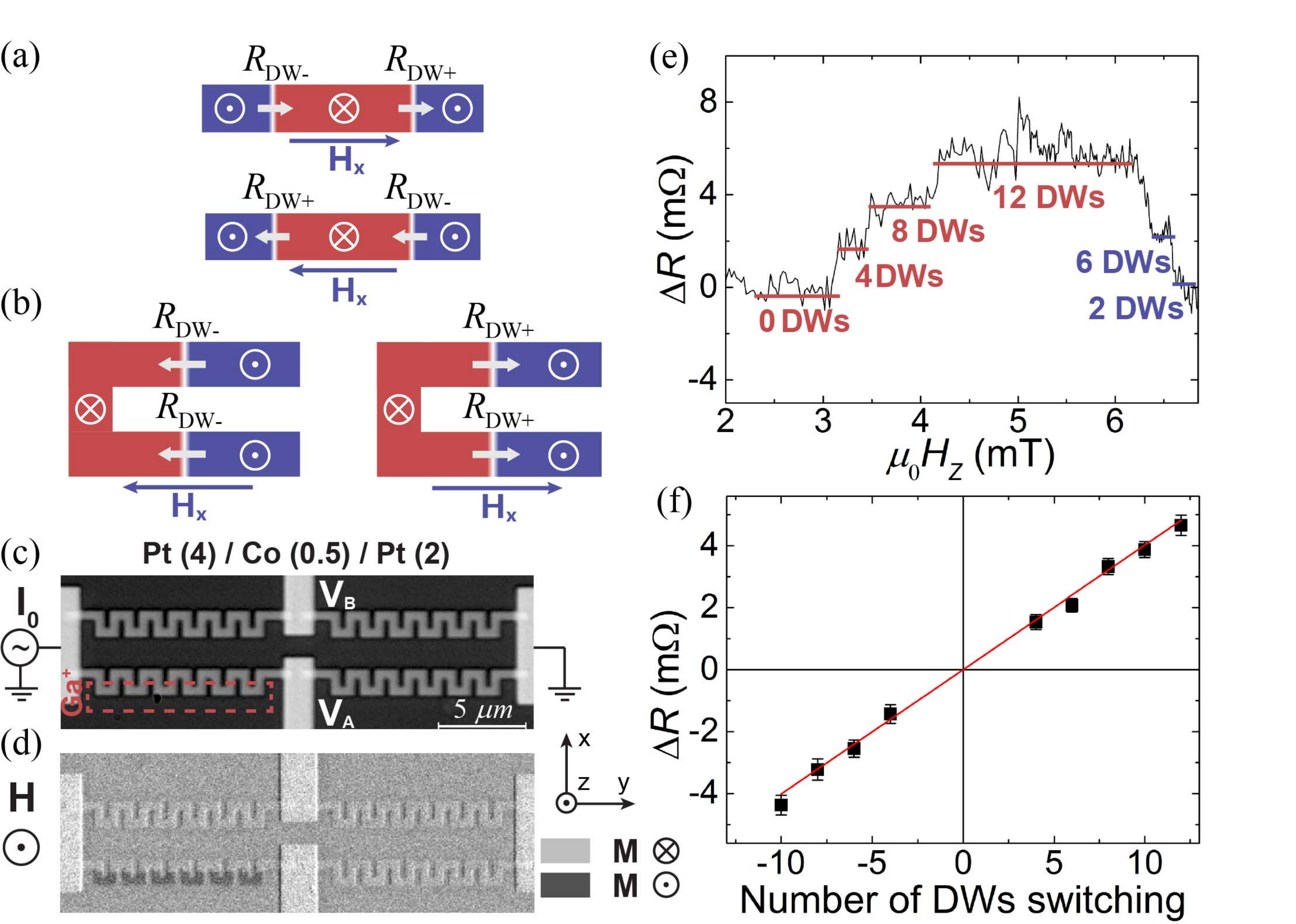}
\caption{(a)(b) A straight (zigzag) magnetic wire where two DWs with opposite (same) chirality are created. (c) Kerr microscopy image of 4 Pt/Co/Pt strips in a Wheatstone bridge configuration, where one of
the strips has been patterned using a focused Ga ion beam. (d) Kerr microscopy image upon applying a perpendicular magnetic field,
showing the nucleation of DWs. (e) Resistance change in the wire as a function of perpendicular field. (f) Magnitude of resistance change as a function of number of DWs nucleating ($+$) or annihilating ($-$). The error bars are single standard deviations. The red solid line is a linear fit.}
\label{Figure3}
\vspace{-1.5em}
\end{figure}

To possibly measure $R_{\textrm{chiral}}$, one could naively use a straight Pt/Co/Pt nanowire as reported earlier \cite{Franken2012c};
see Fig.~\ref{Figure3}(a). Here, the red part indicates the region irradiated by Ga ions which reduce the perpendicular magnetic anisotropy and
allow the nucleation of DWs on two sides \cite{Franken2011}. However, this straight wire is not suitable for measuring the chiral
resistance, as we will experimentally show later on, because the DW pair has an opposite chirality when applying a high in-plane field. Thus, the total chiral resistance $R_{\textrm{DW}+}+R_{\textrm{DW}-}$ is averaged out,
as shown in the inset of Fig.~\ref{Figure2}. In order to overcome this difficulty, we develop an
alternative, zigzag strip [Fig.~\ref{Figure3}(b)]. The zigzag together with anisotropy modifications allows for two DWs with the same chirality where
$C_\textrm{m}$ is controllable by the in-plane field.

Figure~\ref{Figure3}(c) and \ref{Figure3}(d) show Kerr microscope images of the sample structure used for the chiral DW resistance measurements. A perpendicularly
magnetized stack Pt(4 nm)/Co(0.5 nm)/Pt(2 nm) is employed to provide a sizable DMI \cite{Je2013}. Here, the red dashed area indicates the Ga$^+$ irradiation region with a dose of
$0.5\times 10^{13}$ cm$^{-2}$ at 30 keV. Four $1~\mu$m wide zigzag wires are patterned using electron-beam lithography and
a lift-off, in the form of a Wheatstone bridge to improve the signal-to-noise ratio. We apply an AC current with a constant strength $I_\textrm{RMS}=0.25$ mA at a frequency of $f=501$ Hz, and measure the differential
output signal $V_{\textrm{diff}}=V_{\textrm{A}}-V_{\textrm{B}}$ with a lock-in amplifier [See Fig.~\ref{Figure3}(c) for $V_{\textrm{A}}$ and $V_{\textrm{B}}$], which can be used to calculate the change of the resistance in the wire due to the DWs.

Figure~\ref{Figure3}(e) shows the resistance change $\Delta R$ in the wire for DWs whose number is controllable. In the measurement, the magnetization is first saturated by applying $\mu_0
H_z=-10$ mT, then a positive magnetic field is swept from 0 mT to 10 mT. A stepwise increase of $\Delta R$ is observed,
corresponding to the appearance of DWs. The coercive fields of the irradiated domains show small statistical variations so that not all DWs
appear simultaneously. The number of DWs nucleated can
easily be determined by Kerr microscopy (see Supplementary Material). The solid red and blue lines indicate ranges where the number of DWs is constant, and successive increases (decreases) of DWs is indicated by red (blue). In Fig.~\ref{Figure3}(f), $\Delta R$ is plotted as function of the number of DWs present, showing a linear dependence as expected. From the linear fit, we extract a resistance change of $0.38 \pm 0.02 \textrm{ m}\Omega$ per DW, which is comparable to our previous report \cite{Franken2012c}.

The procedure to measure the chiral DW resistance is as follows: (i) The field $H_x$ at which we want to measure $R_\textrm{DW}$ is set, (ii) we saturate the sample at $\mu_0H_z=-10$ mT, (iii) a field of $\mu_0H_z= +1$ mT is applied and $V'_{\textrm{diff}}$ is measured using a lock-in amplifier, (iv) we carefully increase $\mu_0H_z$ to nucleate a few domains, (v) $\mu_0H_z$ is then reduced back to $\mu_0H_z= +1$ mT and $V''_{\textrm{diff}}$ is measured, and the voltage difference $\Delta V=V''_{\textrm{diff}}-V'_{\textrm{diff}}$ is recorded. Steps (iii)-(v) are then repeated to systematically track the resistance change as a function of number of DW's giving an accurate measure of $R_\textrm{DW}$ at a given $H_x$. This procedure is then repeated for every $H_x$. In order to quantitatively compare the measured $R_\textrm{DW}$ as a function of $H_x$ with the theoretical model we should consider the DW resistance in the magnetic Co layer, which can be obtained from $\Delta R$ by using a Fuchs-Sondheimer model \cite{Franken2011} to exclude current shunting through the Pt layers.

\begin{figure}
\includegraphics[scale=0.44]{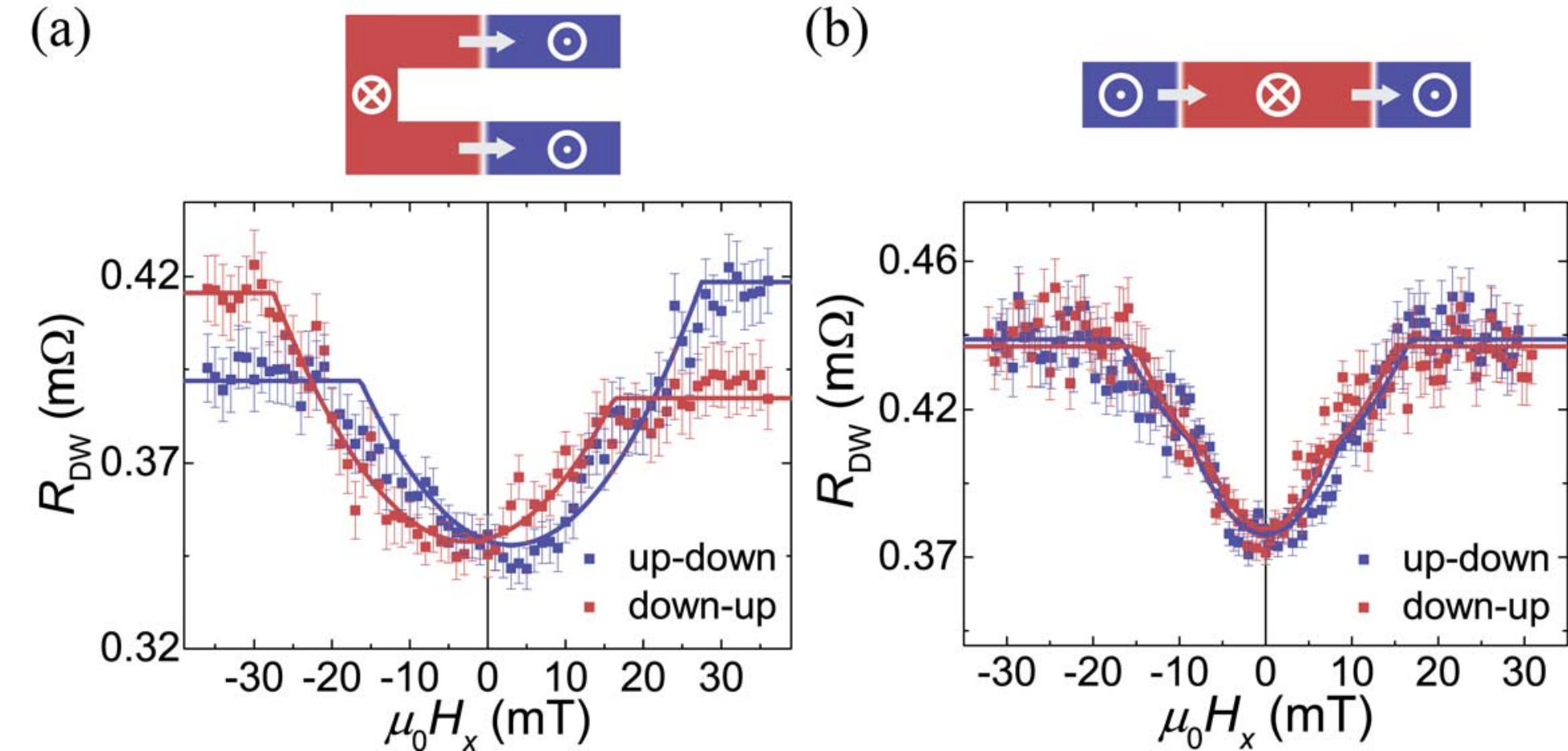}
\caption{The measured DW resistance of a single DW as a function of in-plane field on (a) a zigzag wire and (b) a straight wire. The blue and red squares represent the
magnetic switching with different polarities (up-down and down-up, respectively). The blue and red solid lines are fits to the data based
on the theoretical model. Error bars are single standard deviations.}
\label{Figure4}
\vspace{-1.5em}
\end{figure}

The measured $R_{\textrm{DW}}$ of the Pt/Co/Pt zigzag and straight wires as a function of in-plane field are shown in Figs.~\ref{Figure4}(a) and \ref{Figure4}(b), respectively. The blue and red squares represent the
magnetic switching with different polarities (up-down and down-up). For the zigzag wire, we observe that
$R_{\textrm{DW}}$ rises as $H_x$ increases, owing to an increase of AMR when a Bloch wall transforms into a N\'eel wall. In addition, we clearly observe a field shift, which originates mainly from the DMI \cite{Je2013,Franken2014a,Emori2014} and partially from the chiral MR (see Supplementary Material). As expected from the chirality of the DW spin structures, the field shift has an opposite sign for DWs of opposite polarity. Another prominent observation in Fig.~\ref{Figure4}(a) is that the magnitude of the DW resistance depends on the chirality of N\'eel walls, which can be varied by reversing the in-plane field or changing the polarity of the N\'eel walls. As can be seen, a
substantial difference in resistance at saturation is observed in the zigzag wire, as expected from our theoretical prediction.
Such difference in resistance for opposite chirality of N\'eel walls is strongly supported by the results from the straight wire, where all
chirality-induced effects should be canceled out due to the opposite chirality of two DWs under an in-plane field. Indeed, in contrast
to the results from the zigzag wire, no significant difference between the up-down and down-up data sets are seen in
Fig.~\ref{Figure4}(b). This is additionally confirmed by performing a fit to the data as further explained in Supplementary Material, showing a good agreement with the behavior illustrated in the inset of Fig.~\ref{Figure2}. Consequently, we can conclude that the
aforemeasured resistance difference results from the chirality of DWs.

Next, we estimate the strength of the Rashba effect and DMI by fitting to our theoretical model [Eq.~(\ref{Eq2})]. We assume standard material parameters for Co as used in our previous works \cite{franken2011domain,Franken2014a}:
$A=16\times10^{-12}$ J/m, $M_\textrm{S}=1070$ kA/m, effective anisotropy $K_{\textrm{eff}}=0.28$ MJ/$\textrm{m}^3$, and $\lambda=\sqrt{A/K_{\textrm{eff}}}=8$ nm. The fits are plotted in Fig.~\ref{Figure4}(a) as blue and red lines, which is in a good agreement with the experimental data. Since we fit these two
data sets separately, we extract the free parameters by taking the average of the up-down and down-up data: $\mu_0 H_{\textrm{d}}=34.7$ mT, $\Delta R_{\textrm{AMR}}=50 \textrm{ m}\Omega$,
$\rho_\textrm{DW}=1.41\times10^{-9} \Omega\textrm{ m}$, $D=-0.049 \textrm{ mJ/m}^{2}$, $\mu_0 H_{\textrm{DMI}}=D/\lambda M_\textrm{S}=-5.9$ mT, and
$k_\textrm{R}=-1.43\times10^6 \textrm{ m}^{-1}$ (the parameter errors are presented in Supplementary Material). The AMR parameter is similar to the value $\Delta R_{\textrm{AMR}}=86 \textrm{ m} \Omega$
obtained from a micromagnetic simulation \cite{manago2009magneto,hassel2011domain} (see Supplementary Material), demonstrating that the increase of DW
resistance at small field is indeed due to AMR. The estimated $D$ is comparable to the previously reported value $D=0.1 \textrm{ mJ/m}^{2}$ as measured in a similar Pt/Co/Pt stack \cite{Kobs2011}. For the Rashba coefficient, a first-principles calculation predicts
$k_\textrm{R}=14.9 \times 10^6  \textrm{ m}^{-1}$ for a Pt/Co interface. This value is not representative for our experimental system since a Pt/Co/Pt stack is used where two opposing Pt/Co and Co/Pt interfaces need to be taken into account leading to much smaller effective Rashba coefficient \cite{Kim2013,Freimuth2014,Lavrijsen2012}. Indeed, we extract a $k_\textrm{R}$ which only amounts up to 9.6 \% of the quoted theoretical value.

Since we can independently measure the strength of the Rashba effect and DMI, sharing common origins (SOC and structural inversion asymmetry) in the same system, we can explore their correlation possibly shedding light on their entangled role in SOC-related phenomena
\cite{Kobs2011,Nakayama2013}. For example, it is shown that the Rashba model implies the interfacial DMI energy density $D$ is proportional to a Rashba coefficient in a form of \cite{Kim2013}
\begin{equation}
 D=2 k_\textrm{R} A.
 \label{Eq4}
\end{equation}
Using the values obtained from our fitting, we find $A=D/2k_\textrm{R}= 17\times10^{-12}$  J/m, which is of the same order of magnitude as the exchange stiffness of bulk Co \cite{franken2011domain}. Systematic errors of $D$ and $k_\textrm{R}$ could arise from the uncertainty in $K_\textrm{eff}$ and $\lambda$ due to the Ga irradiation, which we do not separately address in the current work. However, since $D$ and $k_\textrm{R}$ both scales with $\lambda$, the errors do not affect the verification of Eq.~(\ref{Eq4}). This agreement found from the Rashba model and using our experimentally extracted parameters implies that the DMI in our sample may originate from the Rashba-induced twisted Ruderman-Kittel-Kasuya-Yosida interaction \cite{fert1980role,Kim2013,Pyatakov2014,imamura2004twisted} rather than the original explanation by Moriya \cite{Moriya1960} based on a superexchange model. Quantitatively, the Rashba strength is proportional to the ratio of symmetric and antisymmetric exchange.
It implies that the DMI may be tuned by controlling the strength of the Rashba effect (e.g., by applying a electric
field \cite{Liu2011,Barnes2014}) or the strength of the symmetric exchange. In passing, we note that a one-to-one correspondence between the DMI and
symmetric exchange was demonstrated very recently using an optical spin-wave spectroscopy \cite{Nembach2015}.

Although we used the Rashba model to analyze the chiral DW resistance in view of chiral precession of conduction electron spins, another possible explanation might be found in considering the bulk spin Hall effect (SHE) of Pt. As far as symmetry constraints are concerned, it is not impossible for the SHE to generate the chiral DW resistance, though such theory is not developed yet. Specifically, considering that the SHE is of bulk origin, we do not understand how the chiral DW resistance due to the SHE could match with the DMI, which, in our system, is of interfacial origin. In contrast, the good match between $R_{\textrm{chiral}}/R_{\textrm{LZ}}$ and $(\pi\lambda/2)(D/A)$ in our experiment suggests that the chiral DW resistance and the DMI share the common origin.

In summary, we measured the DW resistance on a magnetic zigzag wire and found that it was chirality dependent. This measurement allows us to unambiguously determine the magnitude of
the Rashba effect and DMI as independent parameters. A quantitative agreement was found between experiment and theory, supporting the idea that the Rashba effect could
be coupled to DMI via the symmetric exchange interaction. Besides its fundamental importance, the chiral DW resistance opens up possibilities of designing energy-efficient magnetic DW devices, for example, if we combine it with recently proposed electric-field control of DW chirality \cite{Chen2014}.

\section*{Supplementary Material}
See Supplementary Material for 1. Derivation of chiral DW resistance. 2. DW width variation. 3. Quantification of the DMI by the field shift. 4. The real-time DW resistance measurement. 5. Determination of the anisotropic magnetoresistance. 6. Straight Pt/Co/Pt wire: fitting the experimental data. 7. DW resistance measurement of other samples. 8. DW resistance measurement of Pt/Co/AlOx.
\section*{Acknowledgement}
The authors acknowledge P. Haney, C.-Y. You, M.~D.~Stiles, and J. McClelland for critical reading of the manuscript. This work is part of the research programme of the Foundation for Fundamental Research on Matter (FOM), which is part of the Netherlands
Organisation for Scientific Research (NWO). K.W.K. acknowledges support under the Cooperative Research Agreement between the University of Maryland and the National Institute of Standards and Technology, Center for Nanoscale Science and Technology, Grant No. 70NANB10H193, through the University of Maryland. K.W.K also acknowledges support by Basic Science Research Program through the National Research Foundation of Korea (NRF) funded by the Ministry of Education (2016R1A6A3A03008831). H.W.L was supported by the National Research Foundation of Korea (NRF)
(Grants No. 2011-0030046 and No. 2013R1A2A2A05006237). K.J.L was supported by the NRF
(Grants No. 2011-0027905 and No. NRF-2015M3D1A1070465).
\\
\bibliographystyle{apsrev4-1}
\bibliography{ChiralDWR}

\end{document}